\documentclass[11pt]{article}

\usepackage[margin=1in]{geometry}
\usepackage{amsfonts,amssymb}
\usepackage{graphicx}
\usepackage{natbib}
\usepackage{mdwlist}
\usepackage[svgnames]{xcolor}
\usepackage{datetime,color}
\usepackage{scalefnt}

\usepackage[rightcaption]{sidecap}
\usepackage[bf,font={sf,small}]{caption}
\usepackage{afterpage,float}

\usepackage{xspace} 

\usepackage{enumitem,multirow}
\usepackage[colorlinks=true,linkcolor=NavyBlue,urlcolor=NavyBlue,citecolor=NavyBlue]{hyperref}
\usepackage{ulem}
\usepackage{wrapfig}
\usepackage{lmodern}
\usepackage[T1]{fontenc}

\usepackage{fancyhdr}
\renewcommand{\refname}{References}
\renewcommand{\bibsection}{\section{\refname}}
\long\def\symbolfootnote[#1]#2{\begingroup%
\def\thefootnote{\fnsymbol{footnote}}\footnote[#1]{#2}\endgroup}

\usepackage{sectsty}
\sectionfont{\large}
\subsectionfont{\normalsize}
\subsubsectionfont{\small}

\usepackage[compact]{titlesec}
\titlespacing{\section}{0pt}{*1.5}{*0}
\titlespacing{\subsection}{0pt}{*1.2}{*1.2}
\titlespacing{\subsubsection}{0pt}{*1.2}{*1.2}

\usepackage{fancyhdr}
\makeatletter
\def\p@subsection{}
\makeatother

\newcommand{\bv}{Br\"{u}nt-V\"{a}is\"{a}l\"{a}}

\newcommand{\ind}[1]{_{\mathrm{#1}}}

\def\numax{\nu\ind{max}}
\def\Dnu{\Delta\nu}

\def\Teff{T\ind{eff}}

\begin{document}


\pagenumbering{Roman}

\parbox{16 cm}{~\\~}{~\\}

\begin{center} 
\parbox{15 cm}{
\begin{center} {\Huge\sf\bfseries Seismology of Giant Planets} \end{center}
\begin{center} {\large\sf Chapter 14 of the book Extraterrestrial Seismology - Cambridge University Press (2015)} \end{center}
\begin{center} {\sf Submitted on Arxiv on November 6th, 2014} \end{center}
\vspace{0.5cm}
\begin{center} {\sf Patrick Gaulme$^{1,2}$, Beno\^it Mosser$^3$, Fran\c{c}ois-Xavier Schmider$^4$, Tristan Guillot$^4$} \end{center}
{\small
{\sf $^1$ Department of Astronomy, New Mexico State University, P.O. Box 30001, MSC 4500, Las Cruces, NM 88003-8001, USA} \\
{\sf $^2$ Apache Point Observatory, 2001 Apache Point Road, P.O. Box 59, Sunspot, NM 88349, USA}\\
{\sf $^3$ LESIA, CNRS, Universit\'e Pierre et Marie Curie, Universit\'e Denis Diderot, Observatoire de Paris, 92195 Meudon cedex, France}\\
{\sf $^4$ Laboratoire Lagrange, Universit\'e de Nice Sophia Antipolis, UMR 7293, Observatoire de la Cote d'Azur (OCA), Nice, France}
}
}
\end{center}

\parbox{5 cm}{~\\}


\begin{center}
\noindent 
\parbox{15cm}{
\begin{sf}
\begin{center}Abstract\end{center}
 Seismology applied to giant planets could drastically change our understanding of their deep interiors, as it has happened with the Earth, the Sun, and many main-sequence and evolved stars. The study of giant planets' composition is important for understanding both the mechanisms enabling their formation and the origins of planetary systems, in particular our own. Unfortunately, its determination is complicated by the fact that their interior is thought not to be homogeneous, so that spectroscopic determinations of atmospheric abundances are probably not representative of the planet as a whole. Instead, the determination of their composition and structure must rely on indirect measurements and interior models. Giant planets are mostly fluid and convective, which makes their seismology much closer to that of solar-like stars than that of terrestrial planets. Hence, helioseismology techniques naturally transfer to giant planets. In addition, two alternative methods can be used: photometry of the solar light reflected by planetary atmospheres, and ring seismology in the specific case of Saturn. The current decade has been promising thanks to the detection of Jupiter's acoustic oscillations with the ground-based imaging-spectrometer SYMPA and indirect detection of Saturn's f-modes in its rings by the NASA Cassini orbiter. This has motivated new projects of ground-based and space-borne instruments that are under development. In this chapter, we review the science that seismology could help understand about the four giant planets, the instrumental and modeling approaches, and the most recent observational results. 
 \end{sf}
}
\end{center}

\newpage

\scalefont{1.05}

\setcounter{page}{1}

\thispagestyle{empty}
\tableofcontents
\newpage
\renewcommand{\thepage}{1 -- \arabic{page}}

\addtocontents{toc}{\protect\contentsline {section}{\sffamily Chapter Title -- ``Seismology of Giant Planets''}{}{}}

\setcounter{page}{1}


\parindent 0cm
\parskip .25cm



\section{Introduction}
Seismology of giant planets has long been considered as both a potentially powerful tool for probing their interiors and a natural extension of helioseismology. 
Giant planets are mostly fluid and convective, which makes their seismology much closer to that of solar-like stars than that of terrestrial planets. 
For this, we refer the reader to the introduction chapters about helio and asteroseismology for basic concepts and vocabulary.
By being the biggest and closest, Jupiter has attracted most of the efforts in this domain. Theoretical studies have started in the late 1970s  and first observational attempts were undertaken in the late 1980s. So far, the two major results are a clear detection of acoustic oscillations of Jupiter \citep{Gaulme_2011},  and the signature of Saturn f-modes in the rings by the NASA \textit{Cassini} spacecraft \citep{Hedman_Nicholson_2013}. 

This chapter first exposes the theoretical motivations of developing seismology for giant planets, which mainly stands on an inaccurate knowledge of their interiors  (Sect. \ref{interior}). The next sections focus on two crucial points: why seismology can be done on giant planets (\ref{sect_seismo}), and what can it bring in terms of physics (\ref{sect_inverse}). We then present the observation techniques that have been used or envisioned to detect oscillations (Sect. \ref{sect_techniques}), and the main observational results (Sect. \ref{sect_obs}). 

\section{Interior structure}
\label{interior}
\begin{figure}[t]
\begin{center}
\includegraphics[scale=0.53]{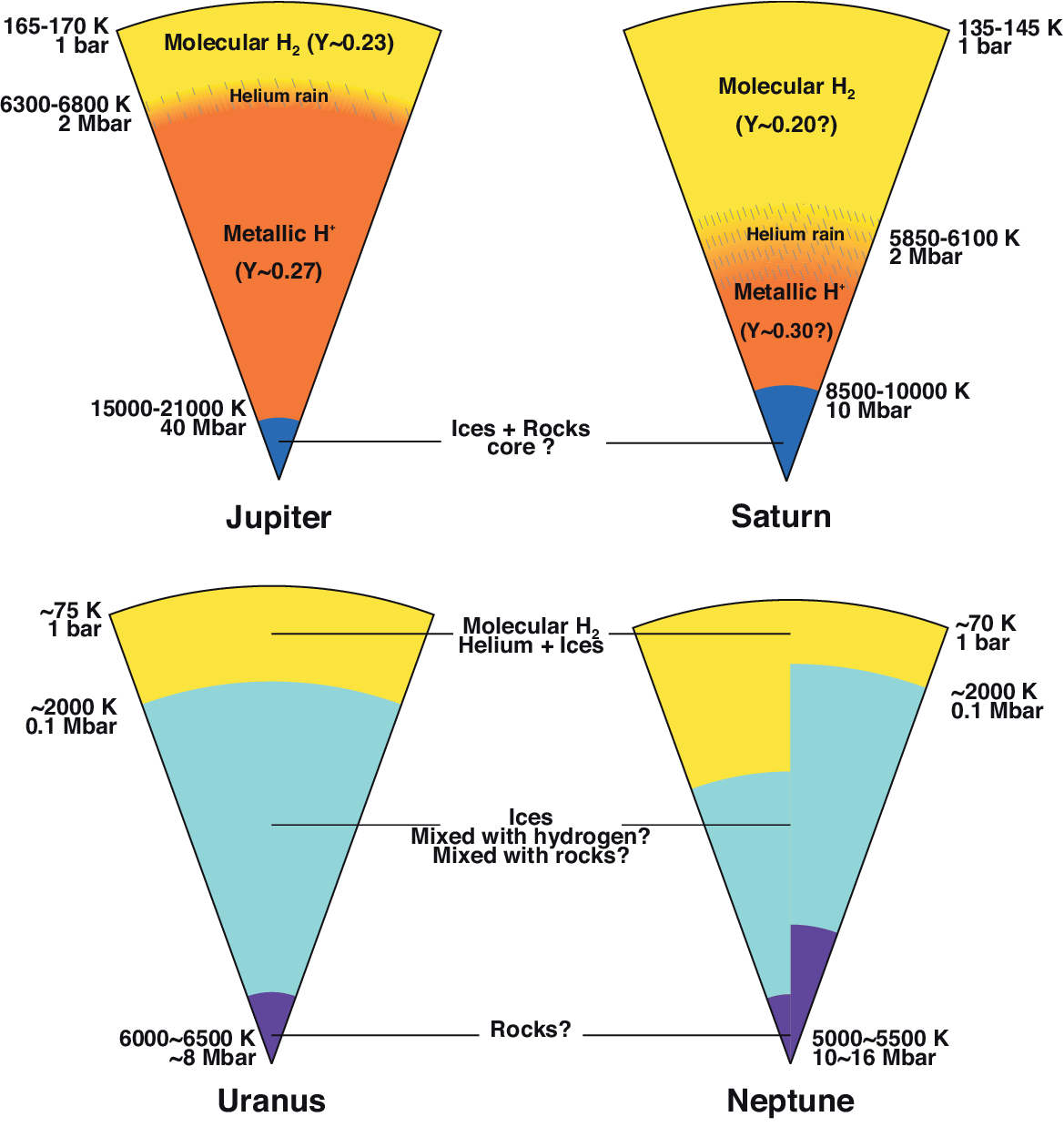}
\caption[Interiors of the four giant planets]
{\textbf{[Colour Figure]} Schematic representation of the interiors of Jupiter, Saturn, Uranus, and Neptune. The range of temperatures for Jupiter and Saturn is for models neglecting the presence of the inhomogeneous region (adapted from \citealt{Guillot_1999}). Helium mass mixing ratios Y are indicated. The size of the central rock and ice cores of Jupiter and Saturn is very uncertain (see text). Similarly, for Uranus and Neptune, considerable uncertainties exist \citep{Nettelmann_2013b}. Recent models of Neptune indicate a wider range of possible solutions of which two representative ones are shown. }
\label{fig_interiors}
\end{center}
\end{figure}

Giant planets are planets massive enough to have retained the hydrogen and helium initially present in the circumstellar disk that led to the formation of the central star and its planets. The study of their composition is important to understand both the mechanisms enabling their formation and the origins of planetary systems, in particular our own. Unfortunately, its determination is complicated by the fact that their interior is thought not to be homogeneous, so that spectroscopic determinations of atmospheric abundances are probably not representative of the planet as a whole. Instead, the determination of their composition and structure must rely on indirect measurements and interior models. Figure \ref{fig_interiors} summarizes the current models of interior structure of the four giants. 

The determination of the fundamental parameters (mass, radius) is a prerequisite, prior to any other characterization, both for the giant planets in our solar system and for giant exoplanets.
The mean density thus provided is a first-order constraint on the mean composition of the planet. Fortunately, for the solar system's giants, a second kind of measurement is available in the form of the departure of their gravity field from sphericity due to their spin. These departures, measured through the planets' gravitational moments, are linked to the interior's density profile and therefore allow, through modeling, for global constraints on the structure and composition \citep[e.g.][]{Zharkov_Trubitsyn_1978}. The uncertainties remain high however. When separating hydrogen and helium from all the other, so-called ``heavy elements'', modern models predict Jupiter has between 10 and 42\,M$_\oplus$ \citep[e.g.][]{Saumon_Guillot_2004, Militzer_2008, Fortney_Nettelmann_2010}, Saturn between 16 and 30\,M$_\oplus$ \citep{Saumon_Guillot_2004, Helled_Guillot_2013,Nettelmann_2013a}, Uranus and Neptune being mostly made of heavy elements, with only 1 to 4\,M$_\oplus$ in hydrogen and helium (\citealt{Podolak_2000}, see also \citealt{Nettelmann_2013b}). Even these uncertain estimates may be changed if semi-convection is widespread in the envelopes of these planets \citep{Leconte_Chabrier_2012}. Part of these heavy elements may be in the form of a central dense core of mass below about 15\,M$_\oplus$ for Jupiter and 20 for Saturn, with no core being an extreme possibility allowed in some cases. 

In 2017, the \textit{Juno} and the \textit{Cassini Solstice} missions will yield much improved determinations of Jupiter's and Saturn's gravity fields, respectively. This should provide ways of better constraining the interiors of these planets. In the case of \textit{Juno}, the combination of gravity, magnetic and radiometric measurements (the latter providing the possibility to constrain the abundance of water in Jupiter's deep atmosphere) will be especially important in providing direct insight into the behavior and properties of Jupiter's deep interior. However, the amount of information carried by the gravitational moments is limited, and that for example only the mean density and second-order gravitational moment $J_2$ are sensitive to the presence of a central core, the higher order moments yielding constraints on layers which are progressively closer to the ``surface''  \citep[][]{Guillot_2005}. The constraints thus brought will remain intrinsically model-dependent. 

Seismic studies of these planets would be a very powerful and complementary way of probing their interiors. As shown for the Sun or the Earth, a carefully determined oscillation spectrum provides a much richer wealth of information on the properties of the interior. It is also very sensitive to discontinuities on the density profiles, which is extremely valuable information to understand the planetary structure. For example, in Jupiter and Saturn, discontinuities (or steep variations) of the density profiles may arise at the core/envelope interface, but also at the location where helium separates from hydrogen and forms droplets, and in double-diffusive interfaces. The discoveries of discontinuities and the determination of their depth would tell us directly which model can be ruled out and provide additional very strong constraints.

\section{Why seismology is possible}
\label{sect_seismo}

The possibility of applying seismology to giant planets is conditional on the existence of a resonant cavity for acoustic waves, and by the presence of a mechanism able to excite acoustic waves.

\subsection{Existence of a resonant cavity}
\label{sect_propag}

\begin{figure}
\includegraphics[scale=0.57]{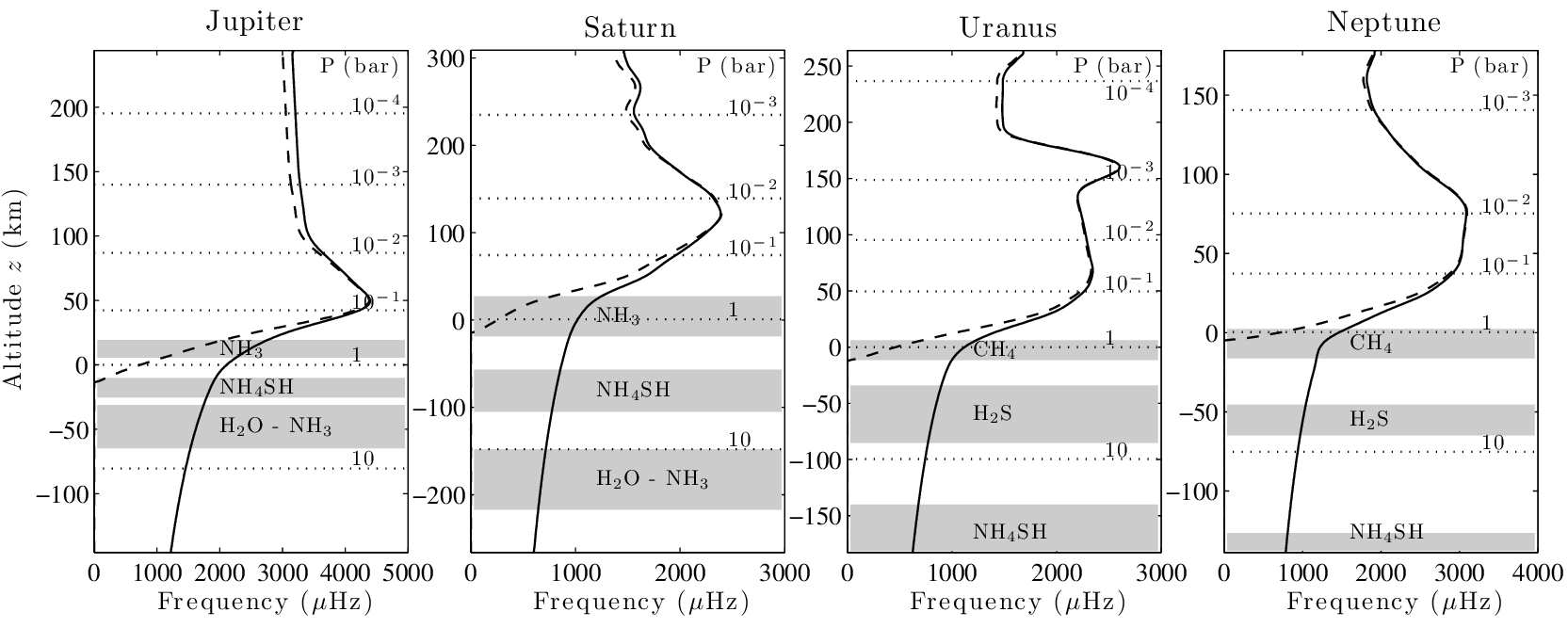}
\caption[Acoustic cut-off and \bv\ frequency profiles]
{Acoustic cut-off $\nu\ind{c}$ (plain line) and \bv\ $N/2\pi$  (dashed line) frequency profiles in upper atmospheres of the four giant planets. Grey regions indicate cloud locations and their chemical compositions \citep[e.g.][]{Irwin_2009}. Above the 1-bar level ($z\equiv0$ km), pressure-temperature profiles come from radio occultation of Voyager spacecrafts \citep[][and references therein]{Lindal_1992}. Below $z=0$, pressure-temperature profiles are adiabats, which imposes $N=0$. }
\label{fig_cutoff}
\end{figure}

The existence of resonant modes of oscillations in a planet requires a trapping mechanism for the waves. As first demonstrated by \citet{Vorontsov_1976}, the atmospheres of the giant planets indeed reflect acoustic waves, but only if their frequency is below a cut-off value, which varies as function of altitude. 
\citet{Mosser_1995} studied the vertical propagation of acoustic waves in Jupiter's atmosphere, with a plane-parallel model, adiabatic propagation, and uniform gravity field. It arises that waves with frequencies shorter than 3\,500\,$\mu$Hz are trapped right below the tropopause level at about the altitude of the top cloud layer, supposedly composed of ammonia ice. The same calculations applied to the three other giants display similar profiles, with waves trapped at the uppermost cloud altitude (Fig. \ref{fig_cutoff}). For Saturn, waves are trapped close to the ammonia cloud decks as for Jupiter. For Uranus and Neptune, even though the extension and thickness of cloud coverage is rather uncertain, waves seem to be trapped near the methane cloud layers.

Besides, from reflectivity factors, \citet{Mosser_1995} put an upper limit on Jupiter's mode lifetime, which is strongly dependent on the mode frequency, from about 300 years at 1\,000~$\mu$Hz to 10 days at 3\,000~$\mu$Hz.
Also, the \bv\ frequency profile is similar to the acoustic cut-off above the tropopause, because both terms vary as $g/H\ind{p}$ - the ratio of local gravity and pressure scale height -, which means that part of the acoustic wave dissipation might transfer into gravity waves in the stratosphere. 


 \subsection{Excitation mechanisms}
 
The confidence that acoustic modes are excited in giant planets mainly relies on the large reservoirs of energy for Jupiter, Saturn and Neptune, because the infrared excess luminosity respectively represents 67, 78, and 161\,\% of the incident solar flux. This inner energy drives a significant convective flux. Uranus stays apart with 6\,\% infrared flux in excess, which indicates that the planet has almost reached the thermal equilibrium.

As known for the Sun \citep[e.g.][]{Goldreich_1994}, convection transfers part of the energy into accoustic waves. But contrary to the case of the Sun for which the forcing is inefficient, \citet{Bercovici_Schubert_1987} show that convective eddies with velocities of $10$ to $80\,\rm m\,s^{-1}$, and length scales on the order of a scale height ($\sim 20$\,km at 1 bar for Jupiter and Saturn)  would have similar periods as those of the standing waves ($\tau_{\rm w}=$4.5 to 9 min) and would thus be well coupled. In that case, they argue that the power per unit volume supplied to the acoustic waves by the turbulent motions would be 
\begin{equation}
W\sim {\rho u^3\over h}M^3,
\end{equation}
where $\rho$ is the density, $u$ and $h$ the typical velocities and scale height of eddies respectively, and $M=u/c\ind{s}$ their Mach number ($c\ind{s}$ is the sound speed). (We neglected a $M^5$ term since $M<1$ is expected). By equating the energy flow from turbulence $Wh$ to that of the waves $F_{\rm w}=\rho c\ind{s} u_0^2$, they obtain the velocity amplitude of the waves: 
\begin{equation}
u_0\sim u^3 / c\ind{s}^2
\end{equation}
By adopting $u\le 80\,\rm$ m\,s$^{-1}$ and $c\ind{s}\sim 1\,$km\,s$^{-1}$, \citet{Bercovici_Schubert_1987} obtain $u_0\le 0.5\,\rm$ m\,s$^{-1}$.

It is not clear at present whether these estimates are realistic. The convective velocities predicted from mixing-length arguments are typically significantly smaller \citep[$u\sim 1\,\rm m\,s^{-1}$ at 1 bar -- see][]{Guillot_2004}. However, values of the vertical velocities obtained in large water clouds typically reach tens of m\,s$^{-1}$ \citep[e.g.][]{Hueso_2002} on the appropriate scale height. Zonal flows also are characterized by velocities over 100\,m\,s$^{-1}$, but this implies much larger (horizontal) scales and it is not clear that they would efficiently couple to the waves.

\subsection{Energy balance and mode lifetime}  

Another approach is to look at the energy balance implied by the presence of $u_0\sim 0.5$\,m\,s$^{-1}$ seismic waves, as implied by the observations of Jupiter \citep[][see Sect. \ref{sympa}]{Gaulme_2011}. In this case, with the above value for $c\ind{s}$ and $\rho\sim 10^{-4}\,\rm g\,cm^{-3}$ in the atmosphere, the energy flux in the waves is $F_{\rm w}\sim 25\,000\rm\, erg\,cm^{-2}\,s^{-1}$. This is to be compared to Jupiter's intrinsic and absorbed solar fluxes, $5\,440$ and $8\,140\rm\,$ erg\,cm$^{-2}$\,s$^{-1}$ \citep[e.g.][and references therein]{Guillot_2005}, implying a maximum available flux $F_{\rm tot}=13\,600\rm\, $erg\,cm$^{-2}$\,s$^{-1}$, respectively. This implies that the energy in the waves must be concentrated spatially, or that there exists a coherent mechanism to excite the waves and that the damping timescale for these is long enough. 

Let us assume that such a mechanism exists and transforms a small fraction $\epsilon$ of Jupiter's energy flux into acoustic waves. In this case the lifetime of the modes is obtained by calculating the time required to channel the energy flux $\epsilon F_{\rm tot}$ into waves of total energy per unit area $\tau_{\rm w} F_{\rm w}$: 
\begin{equation}
\tau_{\rm lifetime}\sim \tau_{\rm w} {F_{\rm w}\over \epsilon F_{\rm tot}}
\end{equation}
If we assume rather arbitrarily $\epsilon\sim 1\%$ (see in a different context \citealt{Showman_Guillot_2002}), then with the above assumptions $\tau_{\rm lifetime}\sim 2$\,days. In this case, the lifetime of the modes is a measure of the damping of the waves which must balance their excitation. Because it is difficult to imagine that the generation of acoustic waves would be more efficient, this implies that the modes should be long-lived (at least for days or more), which is compatible with the quality factor of Jupiter determined by \citet{Mosser_1995}.

\section{What seismology can reveal about interior structures}
\label{sect_inverse}
Given that main-sequence solar-like stars and giant planets are fluid and mostly convective, their oscillation spectra are expected to look similar. They are characterized by a series of peaks that are regularly spaced as function of frequency, whose amplitudes roughly follow a Gaussian shape. The mean separation $\Delta\nu$ in between consecutive mode overtones of a given degree is proportional to the square root of the planetary mean density, whereas the frequency at maximum amplitude $\nu\ind{max}$ is proportional to the acoustic cut-off frequency. It is therefore possible to estimate the ``global seismic parameters'' $\Delta\nu$ and $\nu\ind{max}$ from the asteroseismic scalings \citep[e.g.][]{Kjeldsen_Bedding_1995}. However, two major differences between giant planets 	and solar-like stars that may affect their oscillation spectra are the possible density discontinuities (Sect. \ref{interior}), and the fast rotation, which induces a non-negligible oblateness $e$ \citep{1981AZh....58.1101V,1990Icar...87..198M}. The oblateness turns the mean planetary radius into $R\ind{eq}(1-e/3)$ and thus introduces a relative change of $e/3$ in individual frequencies, $e/2$ on $\Delta\nu$, and $2e/3$ on $\nu\ind{max}$. 
Table \ref{global} presents the global seismic parameters of the four giant planets. The Sun and Jupiter have similar $\Dnu$ since their mean densities are equal, whereas $\numax$ largely differ as do their surface gravities and effective temperatures.  

Two kinds of theoretical studies have been led to fully understand the power of seismology for giant planets. Forward modeling intends to identify relevant features in simulated oscillation spectra computed from planetary models. Inverse modeling intends to quantify how much models are constrained from seismic observations, by inverting simulated oscillation spectra. As regards forward modeling, the oscillation patterns of giant planets have been depicted in many papers \citep[e.g.,][]{Vorontsov_1976,1989SvAL...15..278V, Provost_1993,Gudkova_1995,Jackiewicz_2012}. In particular, \citet{Provost_1993} showed that a density jump at the core boundary of Jupiter would induce a modulation of the regular frequency pattern with amplitude up to $\Dnu/3$ and period of about 10 radial orders, from which it is possible to infer the core size. Also, the influence of upper atmosphere, fast rotation, differential rotation, or chemical phase transitions on individual mode frequencies have been investigated. Overall, comparative works have shown that oscillation patterns based on different planetary models and performed with different seismic codes largely agree \citep[e.g.,][]{Gudkova_1995,Jackiewicz_2012}.


The inverse problem was investigated by fewer authors, and \citet[][]{Jackiewicz_2012} led the most comprehensive work.  From three models of Jupiter's internal structure typical of the current state of the art, they used linear forward modeling to compute the theoretical oscillation spectrum until $l=25$ and build a set of sensitivity kernels, i.e. linear combinations of mode eigenfrequencies and eigenfunctions designed to probe selected regions of the interior. Then, they apply inversion techniques developed for helioseismology to infer details on the expected discontinuities in the profiles of jovian interior. They find that interior boundary layers with sound-speed jump of a few percent could be detected with a sufficient number of observable modes. For deep transitions in the core boundary where fewer modes probe (low $l$), it is very likely that the locations of the large discontinuities could be accurately found, but the magnitudes of the jumps will be more difficult to ascertain because of stochastic noise. They stress that identifying modes until $l = 25$ is important for fine-inverting Jupiter's oscillation spectrum, even for probing the core region. Similar conclusion is expected for the three other giants. 

\begin{table}
 \caption{Comparison of physical and seismic parameters for the four giant planets. Asymptotic $\Dnu\ind{as}$ are obtained as the inverse of the acoustic travel time across the planet, computed from internal structure models (\citet{Saumon_Guillot_2004,Helled_Guillot_2013,Nettelmann_2013b} for Jupiter, Saturn, and Neptune respectively, and Nettelmann priv. comm. for Uranus.). $\Dnu\ind{sl}$ and $\nu\ind{max}$ are extrapolated  from solar values using asteroseismic scalings with correction of planetary oblateness $e$. For Jupiter, observed values from \citet{Gaulme_2011} are shown for comparison.}\label{global}
 \begin{tabular}{lcccccccccc}
   \hline
Planet & $M$ & $R\ind{eq}$ & $e$ & $\Teff$ & $\Dnu\ind{as}$ &$\Dnu\ind{sl}$ & $\Dnu\ind{obs}$ & $\nu\ind{max,sl}$ &  $\nu\ind{max,obs}$ &$\delta\nu\ind{rot}$\\
       &($M\ind{E})$&(km)&(\% )& (K)   &($\mu$Hz) &($\mu$Hz)  &($\mu$Hz)  &($\mu$Hz) &($\mu$Hz)  &($\mu$Hz) \\
\hline
Sun          & $3\,10^5$& 798\,000& 0       & 5777 &       138       &       &  & & 3100  & 0.4\\
\hline
Jupiter     & 317.9        & 71492    &  6.5  & 125    & 152-156  &134 & 155 & 2030 & 1250& 28 \\
Saturn     &  95.1         & 60268    & 10.2 &  95      & 111-115 & 99   &         & 1040 &          & 26 \\
Uranus   & 14.53        & 25559    &  2.4   &  57     &  $\sim172$ & 130 &    &  1030 &          & 17 \\
Neptune & 17.14        & 24764    &  2.7   &  59     & 198-213 & 148 &        &  1270 &          & 19 \\
\hline
\end{tabular}

\end{table}

Beyond pressure oscillations, which have been considered so far because only acoustic waves propagate in convective envelopes, recent asteroseismic observations have shown the huge potential of mixed modes in stars with a radiative core, especially red giants \citep{2011Natur.471..608B}. Such mixed modes result from the coupling of gravity waves in the core with pressure waves in the envelope. The possible presence of a dense and certainly radiative core in Jupiter and Saturn may give raise to such mixed modes: detecting them would provide even finer constraints on the core size, nature and rotation, with respect to pure pressure oscillations.

\section{Techniques to observe acoustic oscillations}
\label{sect_techniques}
Several approaches can be used for detecting acoustic modes: visible photometry for measuring reflected solar flux changes, infrared photometry for temperature fluctuations, and Doppler spectrometry for radial velocities.



\subsection{Visible photometry}
A natural approach is to search for variations of the solar light reflected by Jupiter, resulting from distortions of the planetary external radius $R$ by acoustic modes. \citet{Mosser_1995} shows that a radial mode $l = 0$ of velocity $v$ leads to a variation of the radius $\delta R = c_\mathrm{s}  v/g$, where $c_\mathrm{s}$ is the sound speed in the upper troposphere and $g$ the gravitation. For Jupiter, 
a 1-m~s$^{-1}$ velocity implies $\delta R = 40$ m, with a corresponding variation of reflected luminosity, $(\delta \Phi/\Phi)\ind{refl} = 2 \delta R/R$ of the order of 1 ppm. 

Then, \citet{Gaulme_Mosser_2005} showed that the photometric fluctuations related to planetary radius variations are likely to be negligible with respect to albedo fluctuations of clouds that are caused by acoustic modes on their thermodynamical equilibrium. 
They demonstrated that acoustic waves drive a shift in the gas/condensate phase equilibrium that defines the cloud, which is damped by the kinetics of condensation. The resulting perturbation of the cloud equilibrium can be converted to albedo variations. For acoustic waves of  1-m~s$^{-1}$ amplitudes, photometric fluctuations reach several tens of ppm, which comfortably fits within the photometric performance of current space-based imagers.

Even though additional knowledge on cloud properties of the giant planets is required to assess this result, the photometric approach looks promising for a space mission. Indeed, photometric imaging instruments are much simpler and lighter than high-precision spectrometers and their feasibility and accuracy has been demonstrated by the remarkably successful asteroseismic observations of CoRoT and Kepler \citep{Michel_2008, Borucki_2010}. A first experiment will be realized by the extended Kepler mission K2, which plans to observe Neptune for 3 months with no interruption in early 2015.

\subsection{Thermal infrared photometry}
In the infrared, the photometric detection of global oscillations is based on the approach used in helio- and asteroseismology in the visible domain: acoustic waves are temperature disturbances and lead to fluctuations of brightness temperature.
The black-body emission of most stars peaks in the visible, while it peaks in the mid-infrared for the four giant planets. Thermal emission dominate the reflected solar spectrum at about 6.5 $\mu$m for Jupiter, 10.6 $\mu$m for Saturn, 18 $\mu$m for Uranus, and 15 $\mu$m for Neptune.

The relative Lagrangian temperature perturbation associated with a wave of velocity $v$ expresses as $\delta T/T = (1 - \gamma) v/c_\mathrm{s}$, where $\gamma$ stands for the adiabatic index (e.g. \citealt{Mosser_1995}). For Jupiter, the conversion from wave velocity to temperature is about 0.07 K for 1 m s$^{-1}$. The Lagrangian temperature variation induces a relative flux variation of the black body of
\begin{equation}
\left(\frac{\delta\Phi}{\Phi}\right)\ind{bb} = \frac{hc}{\lambda k_\mathrm{B}T}\ \frac{\delta T}{T}
\end{equation}
Hence, we expect relative photometric variation of about 0.5\,\% for 1 m s$^{-1}$ at 10 $\mu$m, in the case of Jupiter. 

Thermal infrared photometry looks the most favorable technique, because the flux fluctuations are driven by the ratio of wave velocity to the sound speed, instead of the speed of light for Doppler spectrometry (Sect. \ref{sect_doppler_spec}). However, the vertical extension of the atmospheric contribution function in thermal infrared is likely larger than that of visible (limited by clouds), which tends to average out the signature of p-modes, whose vertical wavelengths are about 10 pressure scale-heights in the upper troposphere. In addition, helioseismology has demonstrated that photometric data of solar p-modes are noisier by a factor 10 than Doppler data \citep[e.g.][]{Appourchaux_Grundahl_2013}. Also, detectors are not as sensitive in the mid-infrared as in the visible, and ground-based performance are limited by Earth atmosphere emission. 

\citet{Deming_1989} observed Jupiter in the 8-13 $\mu$m bandwidth, using a 20-element linear array placed parallel to the planet's equator at about 20$^\circ$ in the northern hemisphere. Such an observational set-up totally filtered modes with degrees $l\leq 10$. Observations were performed with the 3-m Infrared Telescope Facility (IRTF) in Hawaii, lasted five days, and did not lead to p-mode detection at the 0.07-K level.

Infrared photometry was then used in attempting to detect acoustic modes caused by the impact of the comet Shoemaker-Levy 9 on Jupiter in 1994 \citep{Lognonne_1994}. \citet{Walter_1996} reported imaging of the planet in a methane emission band at 7.8 $\mu$m with the IRTF, while \citet{Mosser_1996} observed Jupiter with the mid-IR camera of the ESO 3.6-m telescope at La Silla. Both observations did not lead to any detection of impact-excited waves.  Negative results were interpreted a posteriori. Firstly, the wavelength of the waves might have been too short with regard to instrumental resolution, so that their signal was likely averaged out. Secondly, impacts occurred on the night side of Jupiter, so that waves were expected at the limb, where the photon noise is maximum. Thirdly, observations occurred during daytime, which does not favor IR observations. 
Since then, this technique has never been considered anymore neither for ground-based observations nor for space instrument projects. 

\subsection{Doppler spectrometry}
\label{sect_doppler_spec}
The use of Doppler spectrometry for giant-planet seismology was inspired by its success with helioseismology \citep[e.g.][]{Appourchaux_Grundahl_2013}.
The basic principle relies on monitoring the position of a spectral line that probes an atmospheric level where the amplitude of acoustic modes is maximum. 
As for exoplanet search, the minimum spectral resolution to fully resolve spectral lines is about 70\,000. So far, resonant cell and Fourier transform spectrometry have been considered and both allow for spatial resolution.

Atomic resonant cells are used in magneto-optical filters (MOF), whose use for velocity measurements was first studied by \citet{Cacciani_Fofi_1978}. The incoming solar light goes through a cell containing an atomic gas that is present in the solar atmosphere and displays a strong absorption line. The solar atomic line is thermally broadened with respect to that of the atomic cell, which behaves as an extremely thin optical filter. Thus, a shift of the solar line produces a photometric variation of the transmitted light. The instrument sensitivity is the relative flux variation associated with a 1-m\,s$^{-1}$ Doppler shift:
\begin{equation}
S=\ \frac{1}{c}\ \frac{\mathrm{d}\ln I}{\mathrm{d}\ln \lambda} 
\end{equation}
where $\lambda$ is the cell wavelength, $c$ the speed of light, and $I$ the solar spectrum. 

The use of Fourier transform spectrometry (FTS) for Doppler measurements also derives from helioseismology \citep{Scherrer_1995}. The output signal (interferogram) is the Fourier transform of the incoming light spectrum. The presence of deep and evenly spaced absorption lines in the incoming spectrum generates interference fringes at large optical path difference (OPD), of which phase is function of the Doppler shift. The output expresses as $\Phi(\Delta)\,\propto1+\mathcal{C} \cos\left[ 2\pi\sigma_0\Delta(1 + v/c)\right]$, where $\mathcal{C}$ is the fringe contrast, $\sigma_0$ the wavenumber, and $\Delta$ the OPD. The instrument sensitivity is given by the phase shift of an interference fringe per a 1-m\,s$^{-1}$ radial velocity:
\begin{equation}
S =\frac{2\pi\sigma_0\Delta}{c}
\end{equation}

 In a MOF, all photons from a typical 0.01-nm photometric bandwidth contribute to the signal. In a FTS, the filter bandwidth is generally larger (1 to 10 nm), but only $\mathcal{C}/\sqrt{2}\,N$ contribute to the signal. 
When observations are limited by the photon noise only, the amplitude of a velocity signal that is detectable at the 1-$\sigma$ level, is $\delta v=(1/S)(1/$SNR$)$, i.e.
\begin{equation}
\delta v =\frac{1}{S\sqrt{N}}\ \ \mbox{for a MOF, and}\ \ \delta v=\frac{\sqrt{2}}{\mathcal{C}} \frac{1}{S\sqrt{N}}\ \ \mbox{for a FTS.}
\label{eq_photon_noise}
\end{equation}
Therefore, a high performance seismometer must optimize the product of the sensitivity by the square root of the bandwidth.
 
Both types of instruments are sensitive to temperature changes. In a MOF, the reference wavelength is intrinsically stable, but not its transmission. In a FTS, the OPD is temperature dependent. Thermal effects can be minimized by working at a temperature close to an extremum of the sensitivity.
The major drawback of a MOF is that its sensitivity is function of the Doppler shift, making it not uniformly sensitive on a spinning object. The typical sensitivity range of a solar atomic line is $\pm10$ km\,s$^{-1}$. The rotational velocities of Jupiter and Saturn vary of 25.1 and 19.5 km\,s$^{-1}$ along the equator, so that the shift of a reflected solar line is larger than its own width, making a MOF sensitive to a fraction of the planetary surface only. 
For Uranus and Neptune, the whole surfaces can be mapped since their rotational velocity variations along the equator are 5.2 and 5.4 km\,s$^{-1}$ respectively.  Note that the Doppler shift of reflected solar-lines is enhanced by the factor ($1 + \cos\alpha$) where $\alpha$ is the phase angle. 
As we are about see, Doppler spectrometry is the only technique that has led to the detection of Jupiter's oscillations. 


\section{Observational results}
\label{sect_obs}

\subsection{Preliminary detections of Jupiter oscillations}
In 1987, \citet{schmider_1991} conducted full disc observations of Jupiter with a sodium cell MOF mounted on the 1.5-m telescope of the Observatoire de Haute Provence (OHP) providing 7 days of observations with a 23\,\% duty cycle. The data analysis exhibited a periodic signal of Jovian origin, as the signature of the Jovian rotation was detected, between 1000 and 2200~$\mu$Hz. However, the absence of spatial resolution and direct amplitude calibration did not permit to attribute clearly this signal to acoustic modes. 
Later, \citet{Cacciani_2001} designed a double-cell MOF with imaging capability. They tested it on Jupiter, but its actual performance appeared significantly lower than expected, and no Jovian signal could be detected.

In 1991 and 1996, observations were conducted with the Fourier transform spectrometer of the the 3.6-m Canada France Hawaii Telescope (CFHT) \citep{Maillard_Michel_1982}, which was specially set at a fixed OPD, corresponding to a methane absorption band at 1.1 $\mu$m \citep{Mosser_1993, Mosser_2000}. Two circular regions of 10-arcsec diameter, symmetrically located around Jupiter's equator, were sequentially observed to get rid of drifts of the instrument and spurious velocity signal due to Jovian rotation. These observations confirmed the existence of velocity signal from Jovian origin, in the same frequency range as previous observations and with amplitude compatible with theoretical expectations. 

In both MOF and FTS detections, a tentative comb-like structure with a $139 \pm 3\ \mu$Hz mean spacing was also identified. This was probably an artifact due to poor SNR.  This frequency corresponds to the least common multiple of terrestrial and Jovian rotation frequencies, and is incompatible with theoretical estimates of $\Delta\nu$. 
The main lessons of these observations is the need of having spatial resolution. Firstly, it avoids the broadening of solar spectral lines by the fast planetary rotation, which drastically alters the velocity sensitivity. Secondly, analyzing Doppler images by applying masks corresponding to each acoustic mode allows the detection of higher-degree modes.
Thirdly, the comparison of photometric images with Doppler maps permits to get rid of photometric contamination in the velocity signal. These arguments motivated the SYMPA project.

\begin{figure}[t]
\begin{center}
\includegraphics[scale=0.65]{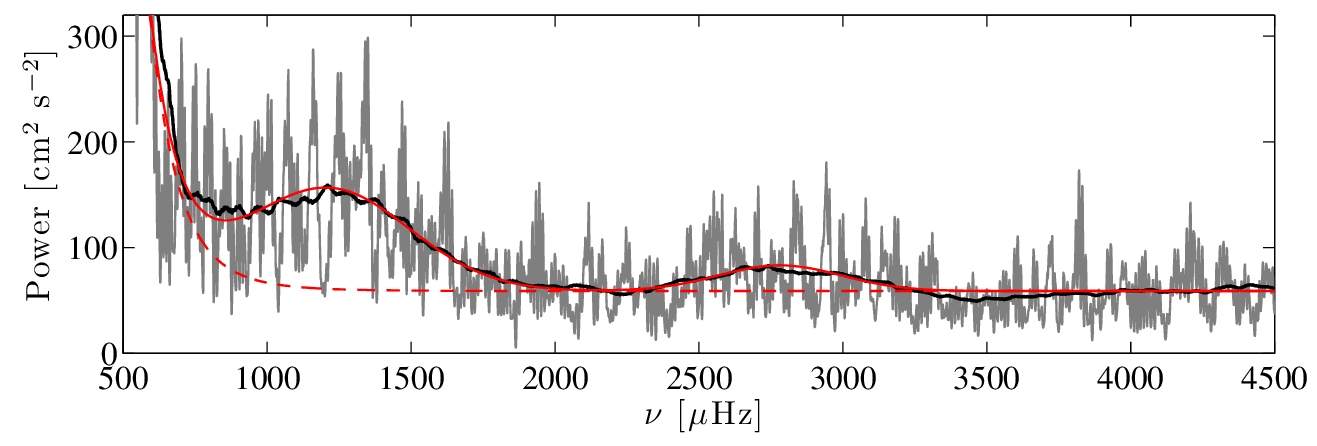}
\caption[Jupiter oscillations as observed with SYMPA]
{\textbf{[Colour Figure]} Power spectrum of the dataset obtained with the SYMPA instrument on Jupiter in 2005 at Teide Observatory
(from Gaulme et al. 2011). The oscillations of Jupiter are characterized both by the increase in power between 800 and 3500 $\mu$Hz, and the regular spacing of the modes ($\Delta \nu=155\,\mu$Hz). The decrease of the mode amplitude around 2000 $\mu$Hz suggests that the propagation of acoustic modes in the upper atmosphere may be a function of frequency.}
\label{fig_sympa}
\end{center}
\end{figure}

\subsection{Detection of Jupiter's global modes with SYMPA}
\label{sympa}
The SYMPA instrument is a Fourier-transform imaging-spectrometer with fixed OPD, whose main optical device is a Mach-Zehnder interferometer \citep{Schmider_2007}. It produces four interferograms of the solar spectrum reflected by Jupiter, in a 5-nm spectral range centered around the Mg triplet at 517 nm. Each interferogram modulates the Jovian photometric figure. Both photometric and interferometric maps are extracted and disentangled from a linear combination of the four images. Next, the velocity maps are decomposed into spherical harmonics
to create a set of time series. The oscillations are sought after in power spectrum of each time series. 

Two SYMPA instruments were built and simultaneously used during two observation runs of Jupiter in 2004 and 2005 at San Pedro Mart\'ir and Teide observatories. The best observation sequence was acquired as part of the 10-day run performed at Teide in April 2005. The data were found to be of much higher quality than all previous observations: at the end of the run, with a 21.5\,\% duty cycle, the mean noise level reached 12 cm s$^{-1}$, which was five times lower than previously achieved \citep{Mosser_2000}. 
However, instrumental imperfections analyzed by \citet{Gaulme_2008} imposed to restrain the search of oscillation modes to low-degrees with antisymmetric pattern. \citet{Gaulme_2011} focused on the time series with best signal-to-noise ratio, which corresponded to modes of degree $l=1$ with significant leakage from $l=2,3,$ and 4.
As in previous Doppler measurements, the time series exhibited excess power between 800 and 2000 $\mu$Hz and secondary excess between 2400 and 3400 $\mu$Hz, but this time the power spectrum was modulated by a clear, high signal-to-noise comb-like structure (Fig. \ref{fig_sympa}). The frequency of maximum amplitude of the main excess power was measured at $\nu\ind{max}=1213\pm50\ \mu$Hz and the mean large spacing at $\Delta\nu=155.3\pm2.2\ \mu$Hz. By assuming that modes are coherent along the observation run, the maximum mode amplitude is $49_{-10}^{+8}$~cm s$^{-1}$. 

The large error bar in the peak frequencies ($\pm7\ \mu$Hz) prevented any mode identification. 
However, the peak structure matches the signature of modes of consecutive degrees (e.g. $l=1$ and $l=2$). Thanks to the thermal control and the imaging capability of SYMPA, the authors could rule out that other effects caused the measured signal, as temperature fluctuations in the interferometer, telescope pointing errors, albedo features on Jupiter, and solar $p$-modes reflected by Jupiter's atmosphere. 
The measured $\Delta\nu$ value agrees with the asymptotic value ($151-156\ \mu$Hz), inferred from the sound speed integral from modern models of the interior of Jupiter \citep{Saumon_Guillot_2004}. 
Better constraints will be provided only by the comparison of observed versus theoretical individual mode frequencies. 

\subsection{Detection of Saturn f-modes by \textit{Cassini}}
\label{ring}
As proposed by \citet{Marley_Porco_1993}, ``the rings of Saturn may act as a seismograph, recording gravitational perturbations associated with acoustic oscillations modes of the planet''. Twenty years later, observations of stellar occultations of stars by the rings made with \textit{Cassini} showed that six density-wave structures detected in the C-ring were compatible with resonances due to Saturn f-modes \citep{Hedman_Nicholson_2013}. By comparing the phase in the density wave for the ingress and egress of the occulted star, the authors roughly identified the azimuthal number of the modes responsible of the observed ring structure. The frequency of the associated mode is easily derived from the position of the feature in the ring. However, a multiplet of three waves associated with a given azimuthal number was found, in contradiction with expectations. 

Only a few f-modes have low enough frequencies to excite resonances in the rings. Since f-mode frequencies are mainly dependent on surface gravity, they are sensitive to the structure of the most external layers of the planet, which includes differential rotation as noticed by \citet{Hedman_Nicholson_2013}. They carry little information about the deep internal structure of the planet. However, \citet{Fuller_2014} proposed that a coupling between surface f-modes and spheroidal or toroidal modes of the solid core could cause the observed multiplets. Although, they did not successfully reproduce the multiplet structure, they argue that a comprehensive description of a giant-planet oscillation spectrum should include couplings with solid bodies modes - as for terrestrial planets - as it might provide valuable information about the existence and the structure of the solid core.

Regarding the f-mode amplitudes, \citet{Marley_Porco_1993} expected that displacements less than 1 m could not produce any noticeable waves in the rings. From the \textit{Cassini} observations, Fuller et al (2014) estimated the f-mode amplitude to be 30 cm. Because their frequencies range between 50 and 100 $\mu$Hz, the induced velocity variations are well below 1 mm s$^{-1}$, which is impossible to detect with Doppler measurements. Actually, such low degree f-modes cannot even be detected on the Sun. 

Besides, these results give a hint about the amplitude of Saturn's acoustic modes. \citet{Goldreich_1994} observed the amplitude of solar acoustic modes to be proportional to $\nu^4$, which would mean - if that applies to giant planets - that Saturn's acoustic mode are larger than 30 cm s$^{-1}$ near $\nu\ind{max}$ ($\sim1000\ \mu$Hz, Table \ref{global}). This is coherent with what was measured on Jupiter and should be detectable with Doppler measurements.

\section{Prospects}

Seismology of giant planets is still in its infancy, and the situation reminds one of helioseismology in the late 1970s and asteroseismology in the 1990s. 
After the recent results on Jupiter and Saturn, there is a clear need to improve the observations and the theoretical models. Improvements in the observations could be done from the ground on Jupiter and Saturn, where oscillation modes with amplitude larger than 5 cm\,s$^{-1}$ and degree up to $l= 25$ could be measured. However, such observations would require a network of 2 to 4-m telescopes evenly placed around the world observing continuously for several weeks. 
Uranus and Neptune would require even larger telescopes. Detection attempts might be envisioned from the ground, but only space missions could achieve the full potential if the technique.

The studies led in the frame of the ESA JUICE mission demonstrated the possibility to reach the objectives with a dedicated instrument onboard a spacecraft approaching the planet.  Two instrumental concepts, a MOF and a Doppler imager \citep{Soulat_2012}, have reached a sufficient technology readiness level (TRL) and are available for future missions, even though no such payload was included in the mission. Several solar-system mission projects contemplate seismic observations of Saturn, Uranus, or Neptune. Such missions could achieve continuous observations for global seismology during several weeks or even months, including during flybys of Jupiter or Saturn by a mission aiming at the ice giants. Observations from an orbiter or during a close fly-by would also allow for studying shallow acoustic waves with solar local-seismic techniques \citep[e.g.][]{Gizon_2010}, which give access to the physical conditions just below the surface.

Alternative approaches may be envisioned too. Impact seismology could be performed if an event similar to the SL9 comet would happen. A strong impact (with an energy deposit of $10^{20}$\,J) would produce detectable tsunami waves and could launch waves probing the upper envelope, and a very strong impact ($> 10^{21}$\,J) should excite $p$-waves that probe the entire planet. In 1994, the strongest impact remained unfortunately about 20 times too small to excite detectable waves, but nowaday's new infrared detectors with enhanced sensitivity and smaller pixels would detect such an event. 

In a broader context, \citet{LeBihan_Burrows_2013} examined the possibility of giant exoplanet seismology. This is out reach with any current observation facility. However, observations like the monitoring of molecular bands on the day-side of the hot Jupiter $\tau$-Bootis along its orbit \citep{Brogi_2012} lets us believe that some of the brightest hot Jupiters might be observable with sufficient SNR to detect oscillations, when extremely large telescopes equipped with high-resolution infrared spectrometers become available.
After its success with the Earth, the Sun and many stars, seismology of giant planets is entering a new era, and will transform our understanding of giant-planet physics and our view of the solar-system history.

\newpage
\bibliography{gaulme_et_al_biblio}
\bibliographystyle{plainnat}  


\end{document}